\documentclass[twocolumn,amsmath,amssymb,floatfix,groupedaddress,prl]{revtex4-1}
\usepackage{leading}
\usepackage[latin3]{inputenc}
\usepackage[makeroom]{cancel}
\usepackage{graphicx}
\usepackage{amsmath}
\usepackage{amsfonts}
\usepackage{amssymb}
\usepackage{color}
\usepackage[left=2cm,right=2cm,top=2cm,bottom=2cm]{geometry}
\usepackage{graphicx} 
\usepackage{dcolumn}
\usepackage{bm}
\usepackage{simplewick}
\usepackage{array}
\usepackage{appendix}
\usepackage{cancel}
\usepackage[normalem]{ulem} 
\RequirePackage[
   hyperindex,colorlinks,bookmarksnumbered,
   plainpages=true,pdfstartview=FitH]{hyperref}
\hypersetup{linkcolor=blue,urlcolor=blue,citecolor=blue}
\usepackage{hyperref}
\usepackage{mathrsfs}

 \renewcommand{\emph}[1]{\textit{#1}}

\definecolor{darkblue}{rgb}{0,0,0.5}
\definecolor{darkgreen}{rgb}{0,0.5,0}
\definecolor{darkred}{rgb}{.7,0,0}
\definecolor{purple}{rgb}{0.5,0,0.6}
\definecolor{orange}{rgb}{1,0.5,0}
\definecolor{grey}{rgb}{.6,.6,.6}
\definecolor{lightpink}{rgb}{1,0.7,0.75}
\definecolor{pink}{rgb}{1,0.4,0.58}
\definecolor{deeppink}{rgb}{1,0.08,0.58}
\usepackage{upgreek}
\begin{document}
\title{Quantum criticality in coupled hybrid metal-semiconductor islands}

\author{D. B. Karki}
\affiliation{Materials Science Division, Argonne National Laboratory, Argonne, Illinois 60439, USA}

\begin{abstract}
We show that the combined effects of dynamical Coulomb blockade and integer quantum Hall effect in a coupled hybrid metal-semiconductor setup provide a pathway for realizing resonant tunneling in Luttinger liquids. This hybrid setup can be brought to the quantum critical regime by varying gate voltages and contact resistances. We explore the nature of quantum criticality, Kondo effect, charge fractionalization and transport in such a hybrid setup, and verify their robust non-Fermi liquid behaviors. Our work opens a promising route for quantum simulating exotic zero temperature quantum critical phenomena associated with Luttinger liquid physics in a nanoengineered electronic circuit with well-defined quantum Hall channels.
\end{abstract}

\maketitle

Quantum impurity models have resulted in remarkable progress towards understanding of quantum criticality in a wide class of phenomena associated with strongly correlated materials~\cite{Wiegmann_1983,Cox_Adv_Phys(47)_1998}. These models usually suggest a simple setting with versatile control to observe appealing features of Fermi-liquid as well as non-Fermi liquid states~\cite{Cox_Adv_Phys(47)_1998}. The most common example of the latter is the Luttinger liquid (LL), which is characterized by a power-law scaling of the tunneling density of states at low energies~\cite{Haldane_1981}. Presence of even a single quantum impurity in LLs drastically changes their transport properties~\cite{Kane_1992,*aa2, exxt}. Quantum impurity problems in LLs have laid great understanding on various paradigmatic one-dimensional phenomena such as Kondo screening, charge and spin fractionalization, and quantum phase transitions~\cite{Giamarchi2003}.

In recent years, the physics of a quantum impurity in LLs has been intensively investigated in more controllable nanoelectronic setups~\cite{fk2,fk3, fk1}. These studies have exploited the known mapping between the problem of one channel conductor in an Ohmic environment and that of the quasiparticle tunneling in a LL~\cite{saffi}. Namely, a hybrid metal-semiconductor nanodevice with an electronic channel realized by a fully tunable quantum point contact (QPC), and a total of $\mathcal{N}$ fully ballistic channels forming a linear series resistance provides an analogous realization of the quantum impurity in LL with an interaction parameter $\eta=\mathcal{N}/\left(\mathcal{N}+1\right)\geq 1/2$~\cite{saffi,fk1}.

When the metal grain used in the setups~\cite{fk2,fk3, fk1} is operated in the regime where the level spacing effectively vanishes, a complete incoherence between the aforementioned electronic channels sets in. Consequently, resonance among all participating electronic channels is lost for $\mathcal{N}>1$. In this case, the gate tunable zero temperature quantum critical point is absent, and the system rather flows toward the intermediate coupling fixed point~\cite{Matveev_1991, *Matveev1995, Furusaki1995b}. The conductance at these fixed points is always smaller than the corresponding unitary value.

The absence of above-mentioned quantum criticality in a single island setup mainly results from the $\mathcal{N}>1$ free ballistic channels. Let's now consider that we have two islands, left and right, each having its own lead (tunable QPC), connected by $\mathcal{N}$ ballistic channels. In this case, the $\mathcal{N}$ free ballistic channels can effectively be fully covered within the left and right islands. The resulting setup then basically works as a single composite island having two tunable QPCs, while the connecting constrictions just provide a dynamical Coulomb blockade environment (with series resistance $R_q=h/e^2\mathcal{N}$) to the remaining tunable QPCs. Therefore, resonant tunneling between two QPCs in an interacting environment can be achieved by tuning the effective gate voltage and contact resistances.

In this work, we show that the mapping~\cite{saffi,fk1} applied to the double hybrid metal-semiconductor islands connected by $\mathcal{N}$ ballistic channels effectively retrieves the resonant feature and hence the Kondo effects initially lacking in the corresponding single island setups. The double island setup could thus form an analogous realization of resonant tunneling in LL with an interaction parameter $\eta=\mathcal{N}/(2\mathcal{N}+1)$. The latter can be verified by simulating the resulting quantum critical point, which supports fractionalized excitations with an effective charge $e^*=\eta e$, and the residual entropy $\ln\sqrt{\eta}$. Moreover, by adjusting the number of islands, discrete sets of $0<\eta\leq 1/2$ can be implemented. In addition, we show that when some of the connecting constrictions deviate from the full ballistics, the couple hybrid system flows toward the intermediate coupling fixed point characterized by the oscillatory behavior of conductance. In the latter case, interaction strength $1/2<\eta<1$ can be realized but without resonant features.  Importantly, special limits of our approach can fully retrieves the physics of widely studied two-channel Kondo effect ($\eta\to 1/2$)~\cite{Furusaki1995b}, and also that of the double charge quantum islands ($\eta\to 1/3$)~\cite{ccd, *KBM1}.

\begin{figure}[t]
\begin{center}
\includegraphics[scale=0.28]{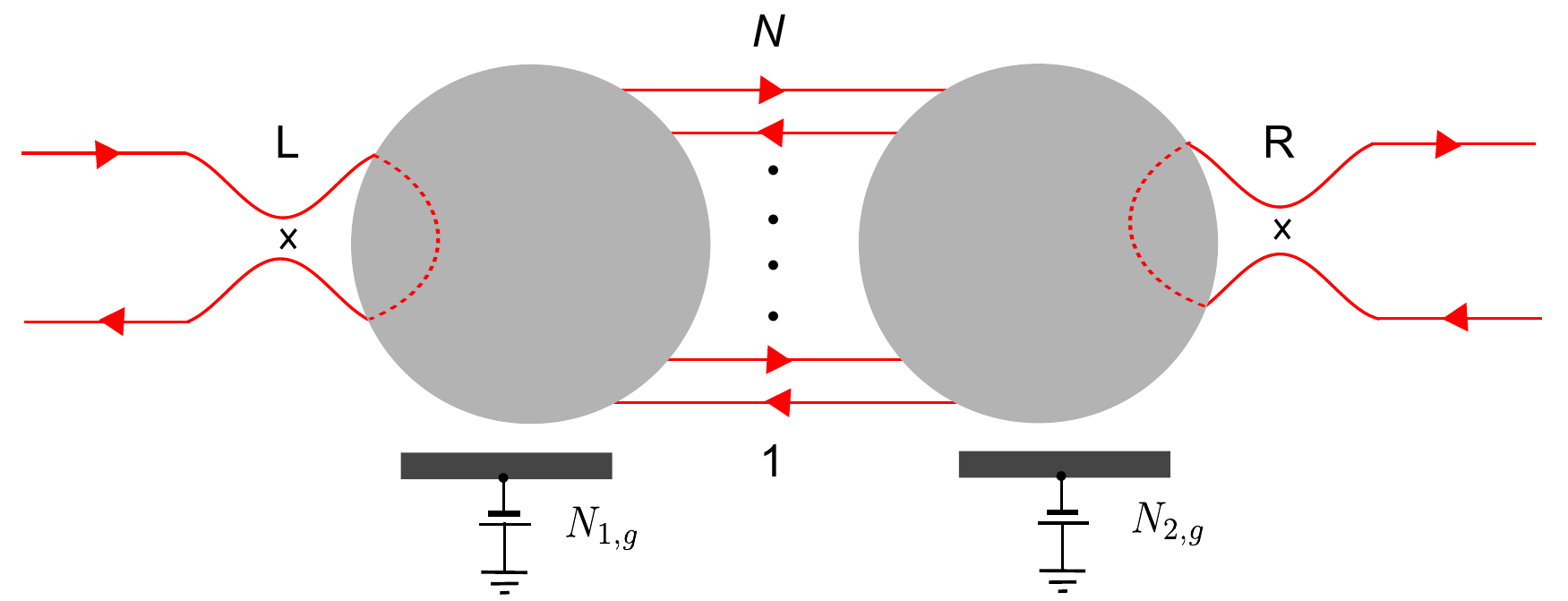}
\caption{Schematic of experimental setup comprising two charge grains each having one fully tunable QPC, and are connected by $\mathcal{N}$ ballistic channels. The $\times$ signs in left and right QPCs represent their backscattering/tunneling centers.}\label{deepak}
\end{center}
\end{figure}

In the following, we make the above articulations more explicit by presenting detailed calculations based on bosonization technique. To this end, we consider a setup comprising double charge islands connected to each other by $\mathcal{N}$ fully ballistic QPCs and two fully tunable QPCs attached to the individual islands as shown in Fig.~\ref{deepak}. We focus on the regime where the charging energy of the grains is the largest, and their level spacing is the smallest energy scale in the problem~\cite{Matveev1995, Furusaki1995b}. The occupation of each island can be tunned by applying voltage $V_{1/2, g}\propto N_{1/2, g}$ to the corresponding external gates. Note that since the connecting constrictions are fully ballistic, only the combination $\tilde{N}_g=(N_{1, g}+N_{2, g})/2$ is sufficient for the description of transport and thermodynamical properties. We also note that when the number of connecting QPCs grows to an infinitely large value, the series resistance gradually vanishes and hence the setup in Fig.~\ref{deepak} is formally identical to the single island case, which also exhibits a two-channel Kondo effect~\cite{Pierre_2015}. As already announced earlier, the case of $\mathcal{N}=1$ would be similar to the recently realized two-site charge Kondo setup~\cite{ccd, *KBM1}.

The spin-polarized electrons in $\mathcal{N}+2$ interspaced QPCs realized in two-dimensiobnal electron gas can be described by the quadratic Hamiltonian of the form~\cite{Wen_1990}
\begin{align}
H_0 &=\frac{v_F}{2\pi}\sum_{i=1}^{\mathcal{N}+2}\int^{\infty}_{-\infty}dx \left[\left(\frac{\partial\varphi_i}{\partial x}\right)^2+\pi^2\Pi^2_i\right],\label{ch0}
\end{align}
where $v_F$ is Fermi velocity, and the bosonic field $\varphi_i$ satisfies the usual commutation relation with its momentum density $\Pi_i$, $\left[\varphi_i(x), \Pi_i(y)\right]=i\delta\left(x-y\right)$. In addition, we assumed $\varphi_{\mathcal{N}+1}=\varphi_L$, and $\varphi_{\mathcal{N}+2}=\varphi_R$. The effects of charging energy in the double island setup can be accounted for by the constant interaction model. Assuming $E^{(1,2)}_c$ be the effective charging energy of the grains 1 and 2, the charging energy Hamiltonian can be written as~\cite{KBM} 
\begin{align}
H_c =&\phantom{-}\frac{E_c^{(1)}}{\pi^2}\left(\varphi_L-\sum_{i=1}^{\mathcal{N}}\varphi_i+\pi N_{1, g}\right)^2\nonumber\\
&+\frac{E_c^{(2)}}{\pi^2}\left(\sum_{i=1}^{\mathcal{N}}\varphi_i-\varphi_R+\pi N_{2, g}\right)^2.\label{che}
\end{align}
In addition to $H_0$ and $H_c$, the Hamiltonian accounting for the weak/strong tunneling from the left and right QPCs needs to be considered. For simplicity, we primarily focus on the strong tunneling regime where weak backscattering in the left and right QPCs can be accounted for by the usual boundary Hamiltonian
\begin{align}
H_b &=\sum_{i=L, R}\frac{\Lambda|r_i|}{\pi}\cos\left[2\varphi_i\left( \tau\right)\right],\label{chb}
\end{align}
where $|r_L|\ll 1$ and $|r_R|\ll 1$ stand for the bare reflection amplitudes of the left and right QPCs respectively, and $\Lambda$ is the high energy cutoff. In the following, we are interested in evaluating the charge current flowing from left to right QPC, which is defined by the operator $\hat{I}=(e/2\pi)\partial_\tau(\varphi_L+\varphi_R)$.

To make further progress, we introduce a new field defined by $\overline{\varphi}=\sum_{i=1}^{\mathcal{N}}\varphi_i/\sqrt{\mathcal{N}}$, and diagonalize the charging energy Hamiltonian~\eqref{che} by introducing three orthogonal bosonic fields $\varphi_j,\;j=a, b, c$~\footnote{$\left(
\begin{array}{ccc}
 \varphi_L \\
 \overline{\varphi} \\
 \varphi_R \\
\end{array}
\right)=\left(
\begin{array}{ccc}
 \sqrt{\eta } & -\frac{1}{\sqrt{2}} & \sqrt{\frac{1}{2} (1-2 \eta )} \\
 \sqrt{1-2 \eta } & 0 & -\sqrt{2 \eta } \\
 \sqrt{\eta } & \frac{1}{\sqrt{2}} & \sqrt{\frac{1}{2} (1-2 \eta )} \\
\end{array}
\right)\left(
\begin{array}{ccc}
 \varphi_a \\
\varphi_b \\
 \varphi_c \\
\end{array}
\right)$}. Assuming $E_c^{(1)}=E_c^{(2)}=E_c$ for simplicity~\footnote{Different charging energies do not affect any of our universal results. When the charging energies are different, the mass terms of the gapped modes $\varphi_{b, c}$ in Eq.~(5) will be renormalized accordingly without affecting the charge mode $\varphi_a$.}, the Hamiltonians~\eqref{ch0}$-$\eqref{chb} then change to the form
\begin{align}
H'_0 &=\frac{v_F}{2\pi}\sum_{j}\int^{\infty}_{-\infty}dx \left[\left(\frac{\partial\varphi_j}{\partial x}\right)^2+\pi^2\Pi^2_j\right],\label{sg1}
\end{align}
\begin{align}
H'_c =&\frac{E_c}{\pi^2}\left(\varphi_b^2+\frac{1}{1-2\eta}\varphi_c^2\right),\label{sg2}
\end{align}
and
\begin{align}
H'_b =&\phantom{-}\frac{\Lambda|r_L|}{\pi}\cos\left[\sqrt{4\eta}\varphi_a{-}\sqrt2\phi_b+\sqrt{2{-}4\eta}\varphi_c{-}4\pi N_L\right]\nonumber\\
&+\frac{\Lambda|r_R|}{\pi}\cos\left[\sqrt{4\eta}\varphi_a{+}\sqrt2\phi_b{+}\sqrt{2{-}4\eta}\varphi_c{+}4\pi N_R\right].\label{sg3}
\end{align}
Similarly, the current operator transforms to the form
 \begin{equation}
\hat{ \mathcal{I}}=\frac{e\sqrt\eta}{\pi}\frac{\partial\varphi_a}{\partial\tau}.\label{rotc}
 \end{equation}
The symbols $N_{L/R}$ in Eq.~\eqref{sg3} are defined as
\begin{align}
N_L &=\frac{1}{2}\left[\left(1-\eta\right)\!N_{1, g}+\eta N_{2, g}\right]\nonumber\\
N_R &=\frac{1}{2}\left[\eta N_{1, g}+(1-\eta) N_{2, g}\right].\label{sg4}
\end{align}
Equations~\eqref{sg3} and~\eqref{sg4} show that only the combination $N_L+N_R=\left(N_{1, g}+N_{2, g}\right)/2= \tilde{N}_g$ is responsible for the description of charge current as already announced earlier. If the connecting constrictions also have a finite reflection coefficient, the gate fluctuation $ \Delta N_g=N_{1, g}-N_{2, g}$ would have crucial consequences of lifting the degeneracy (see below).

In the above Hamiltonians, $\varphi_a$ is the charge mode which acquires the $\eta$ rescaling due to the Coulomb blockade environment formed by connecting ballistic constrictions. The field $\varphi_a$ represents a plasmonic motion which is not affected by the charging energy and causes the dissipation in the system. On the other hand, the modes $\varphi_{b, c}$ acquire a mass gap because of large charging energy, and thus remain neutral for the description of low energy ($<E_c$) charge transport. Further progress can be made by integrating out the gapped modes $\varphi_{b, c}$, which results in an effective Hamiltonian
\begin{align}
H_{\rm eff} =&\phantom{-} \frac{v_F}{2\pi}\int^{\infty}_{-\infty}dx \left[\left(\frac{\partial\varphi_a}{\partial x}\right)^2+\pi^2\Pi^2_a\right]\nonumber\\
&+\left(\frac{e^{\bf{C}}E_c}{\pi \Lambda}\right)^{1/2}\frac{\tilde{r}\Lambda}{\pi}\cos\left(\sqrt{4\eta}\varphi_a+\Theta\right),\label{sg10aa}
\end{align}
where $\Theta$ is an unimportant phase factor, and $\bf{C}$ stands for the Euler's constant. The effective reflection coefficient $\tilde{r}$ is expressed in terms of renormalized reflections and the effective gate voltage as
\begin{align}
\tilde{r} &=\sqrt{|\tilde{r}_L|^2+|\tilde{r}_R|^2+2 |\tilde{r}_L||\tilde{r}_R| \cos\left(4\pi \tilde{N}_g\right)},\nonumber\\
&\qquad|\tilde{r}_{L/R}| =\left(\frac{e^{\bf{C}}E_c}{\pi \Lambda\left(1-2\eta\right)}\right)^{\frac{1}{2}-\eta}|r_{L/R}|.\label{sg8}
\end{align}
Equation~\eqref{sg10aa} represents an analogous realization of resonant tunneling in LL with the interaction parameter $\eta$. Equivalently, Eq.~\eqref{sg10aa} bears similarity with the corresponding Hamiltonian for the single island setup with two QPCs operating on the fractional quantum Hall regime.

The effective reflection coefficient~\eqref{sg8} vanishes for typical choice of parameters $\tilde{N}_g=(2n+1)/4$, and $|\tilde{r}_R|=|\tilde{r}_R|$. This point in the parameter space constitutes the quantum critical point where backscattering processes interfere destructively, giving maximal conductance $G_0=\eta e^2/h$ irrespective of temperature and/or voltage. Away from the critical point $\tilde{r}\neq 0$, the cosine term~\eqref{sg10aa} is a relevant perturbation with scaling dimension $\eta\leq 1/2$. It would thus result in decreasing conductance with decreasing temperature to eventually reach a low-energy fixed point with vanishing conductance. To see these features, in the following, we evaluate the temperature scaling behavior of conductance.

Using the definition of the current operator~\eqref{rotc} in the standard Kubo formula, and treating the backscattering~\eqref{sg3} within the second-order perturbation theory over quadratic Hamiltonian, we obtain the linear conductance
\begin{align}
G=\frac{\eta e^2}{h}\left[1-\mathcal{C}_1\left(\frac{e^{\bf{C}} E_c}{\pi^2 T}\right)^{2-2\eta}\!\!-\mathcal{C}_2\left(\frac{\pi^2 T}{e^{\bf{C}} E_c}\right)^{2\eta}\right],\label{cd1}
\end{align}
where we introduced new notations
\begin{align}
&\mathcal{C}_1 =\frac{\eta }{2}\frac{\sqrt\pi\;\Gamma(\eta)}{\Gamma\left(\eta +\frac{1}{2}\right)}\frac{|r_+|^2}{(1-2\eta)^{1-2\eta}},\nonumber\\
&\mathcal{C}_2 =\frac{\eta e^{2\bf{C}}}{2}\frac{\sqrt\pi\;\Gamma(\eta+1)}{\Gamma\left(\eta +\frac{3}{2}\right)}\frac{|r_-|^2+(1-2\eta)^3|r_+|^2}{(1-2\eta)^{1-2\eta}},\nonumber\\
&r_\pm =|r_L|e^{-2\pi i\tilde{N}_g}\pm |r_R|e^{2\pi i \tilde{N}_g}.
\end{align}
The generality of our result expressed by Eq.~\eqref{cd1} is far-reaching. The widely studied phenomena of the two-channel charge Kondo effect, and the recently studied double-charge Kondo effect are the special limit of our general formula Eq.~\eqref{cd1}. When the connecting constrictions are infinitely large, i.e.,  $\mathcal{N}\to\infty$, the LL parameter takes the value $\eta\to 1/2$. Equation~\eqref{cd1} then retrieves the well known result for the conductance of the single grain setup studied in Ref~.\cite{Furusaki1995b} upon rescaling $E_c\to 2 E_c$, and $\tilde{N}_g\to N_g/2$. For $\mathcal{N}=1$, Eq.~\eqref{cd1} gives the recently established result for the conductance of the double-charge Kondo circuit~\cite{KBM}.

The perturbative expression of conductance~\eqref{cd1} is valid for $T>T^*$, where the crossover temperature is identified as $T^*=e^{\bf{C}}E_c\mathcal{C}_1^{\frac{1}{2-2\eta}}/\pi^2$. At the critical point mentioned earlier, this crossover scale vanishes and the temperature correction to the unitary conductance acquires the scaling behavior $(T/E_c)^{2\eta}$ effectively extending the range of validity of~\eqref{cd1} down to zero temperature. When two QPCs (left and right) are tunned to weak tunneling regime, the charge in the composite
island is nearly quantized (but not in individual islands)
and thus Matveev's charge Kondo mapping could be exploited to the composite island system. In this case the power law describe by the third term of Eq.~\eqref{cd1} would evolve as $\sim (T/T_K)^{2\eta}$ for $T_K$ being the corresponding Kondo temperature.

In the narrow vicinity of critical point defined by the gate voltage and reflections coefficients such that $\tilde{N}_g=\frac{1}{4}+\delta N_g$, and $|r_{R/L}|=r\pm\delta r$, the crossover scale acquires the scaling behaviors $T^*\sim \delta r^{1/(1-\eta)}$, and $T^*\sim \delta N_g^{1/(1-\eta)}$ as a manifestation of LL physics.

It is straightforward to introduce voltage into our problem by gauge shifting of the charge mode $\phi_a(\tau)\to\phi_a(\tau)+\sqrt\eta(eV/2)\tau/\hbar$. The resulting, average current $\left<\hat{\mathcal{I}}\right>$ and shot noise $S$ in the presence of backscattering~\eqref{sg10aa} gives the Fano factor $\mathcal{F}=S/\delta\left<\hat{\mathcal{I}}\right>=\eta$. The latter corresponds to the backscattering of fractional charges $e^*=\eta e$. These fractional excitations belong to the dominant $p=2$ sequence of hierarchal Jain's states $\nu=n/(np+1)$~\cite{jjs}. In addition, making the usual connection of Eq.~\eqref{sg10aa} with the boundary sine-Gordon model, we retrieve the fractional entropy $\ln\sqrt\eta$~\cite{aa1}. We note that by adding more charge grains in an identical way as in Fig.~\ref{deepak} all $\eta\leq 1/2$ can be implemented, such as for three islands $\eta$ would be modified to $\mathcal{N}/(2\mathcal{N}+2)$.

\begin{figure}[b]
\begin{center}
\includegraphics[scale=0.28]{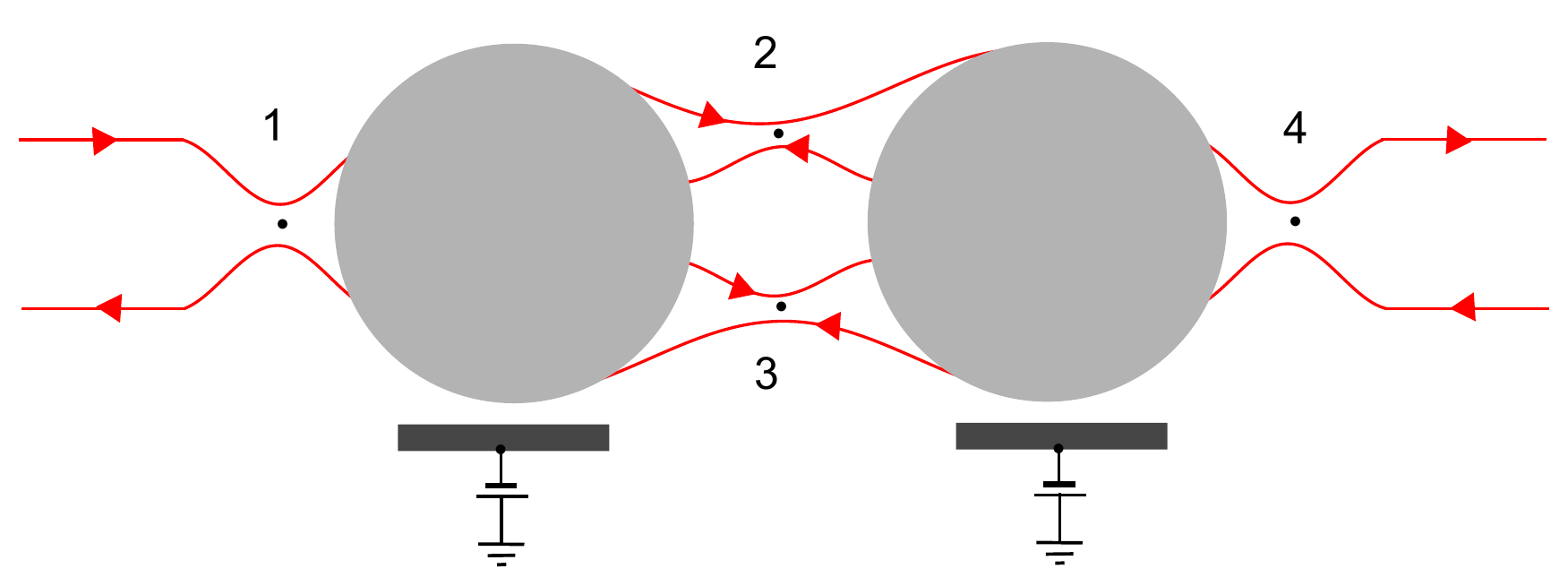}
\caption{Gate tunable double charge grains connected to each other by two QPCs with finite reflections. The dots in QPCs respresnt their backscattering centres. }\label{sfig1}
\end{center}
\end{figure}
In the following, we study the effects of finite reflection on the connecting constrictions. As a minimal setup, we consider just two connecting QPCs along with the two side QPCs as shown in Fig.~\ref{sfig1}. We note that this setup can readily be formed by combining two three-channel charge Kondo (3CK) circuits~\cite{Pierre_2018} in a different way than the recent experimental setup~\cite{ccd} formed by connecting two 3CK circuits such that two islands are singly connected by a QPC.

For simplicity, we assume that the bare reflection coefficients of connecting QPCs are equal $|r_2|=|r_3|=|r_i|$, and similarly to that for the side QPCs $|r_1|=|r_4|=|r_o|$. In addition, we assume the same gate voltage $N_{1, g}=N_{2, g}=N_g$ for both islands. As already announced earlier, since the connecting constrictions have finite reflections, the charge fluctuation between two islands greatly affects the charging effect. Equivalently, the two connecting channels can not be transformed into a single channel, as in the case studied earlier. Nevertheless, it is easy to anticipate the form of low energy effective Hamiltonian describing the transport in the setup~\ref{sfig1}. Namely, since we have four QPCs and two charge grains, the two bosonic modes out of four describing the total Hamiltonian will acquire the gap because of large charging energy. These modes can be integrated out for the description of low energy physics. The remaining two gapless modes, $\varphi_{A, B}$ thus describe the low energy physics of the setup~\ref{sfig1}. It is important to note that these gapless modes are coupled by the interaction (charging energy), and the effective backscattering Hamiltonian acquires a nontrivial form. 

Following the same line of calculations presented earlier, we arrive at the effective Hamiltonian describing the setup shown in Fig.~\ref{sfig1} in the form 
\begin{align}
\mathcal{H}_{\rm eff} &= \frac{v_F}{2\pi}\sum_{i=A, B}\int^{\infty}_{-\infty} dx\left[\left(\frac{\partial\varphi_i}{\partial x}\right)^2+\pi^2\Pi^2_i\right]\nonumber\\
&+\!A\!\cos \left(\!\!\sqrt{\frac{8}{5}} \varphi _A\!\!\right)\!{+}B\!\cos( \sqrt{2} \varphi _B) \cos \left(\!\!\!\sqrt{\frac{2}{5}} \varphi _A\!\!\right).\label{h9}
\end{align}
Constants in Eq.~\eqref{h9} are defined as
\begin{align}
A= &\frac{2\Lambda|r_o| 5^{1/10}}{\pi }\left(\frac{e^{\bf{C}} E_c}{\pi \Lambda}\right)^{3/5}\!\!\cos2\pi N_g\equiv a\Lambda\left(\frac{E_c}{\Lambda}\right)^{3/5},\nonumber\\
B= &\frac{2\Lambda|r_i|5^{2/5}}{\pi } \left(\frac{e^{\bf{C}} E_c}{\pi \Lambda}\right)^{2/5}\equiv b\Lambda\left(\frac{E_c}{\Lambda}\right)^{2/5},\label{ktt}
\end{align}
and operator describing the charge flow from left to right QPC takes the usual form $\hat{\mathcal{I}}=\sqrt{2/5}(e/\pi)\partial_\tau\varphi_A$.

The first term of the backscattering part in the Hamiltonian~\eqref{h9} describes the direct tunneling of a composite fermion while the second term represents the coupling between the two species of such a fermions. It is therefore expected that the strength of the second backscattering term in Eq.~\eqref{h9} is weaker than that of the first term. Indeed, the scaling dimension of first and second terms of the backscattering perturbation in Eq.~\eqref{h9} are $2/5$ and $3/5$ respectively. We note that perturbations in Eq.~\eqref{h9} are competing in the sense that the parameter $A$ is gate tunable and vanishes at $N_g=(2n+1)/4$.

Straightforward evaluation of linear conductance using the second order perterbuation theory results in
\begin{align}
G &=\frac{2 e^2}{5h}\Big[1-\left(\frac{2a^2}{5} \mathcal{J}_{2/5}+ \frac{b^2}{20}\mathcal{J}_{3/5}\right)\Big],\nonumber\\
\mathcal{J}_\alpha &=\frac{\pi^2}{2}\left(\frac{E_c}{\pi T}\right)^{2-2 \alpha}\frac{\sqrt{\pi}\Gamma(\alpha)}{\Gamma\left(\alpha+\frac{1}{2}\right)}.\label{h6}
\end{align}
If both $a$ and $b$ defined in Eq.~\eqref{ktt} are non-zero, the expression of conductance~\eqref{h6} is valid only in the perturbatve regime, i.e., $|r_o|^{5/3}E_c\ll T$, and $|r_i|^{5/2}E_c\ll T$. In this case, the system reaches the intermediate coupling fixed point with competing power law behaviors. It is important to note that the conductance~\eqref{h6} has oscillatory behavior even at the intermediate coupling fixed point. The latter is in contrast to the corresponding conductance behavior of multichannel ($>2$) charge Kondo setups~\cite{Matveev1995}, and is a typical feature of coupled hybrid setups. In addition, since the parameter $a$ depends on the gate voltage, there exists a crossover scale $T^c$ where the two temperature corrections in Eq.~\eqref{h6} become the same order of magnitude. For a given temperature and the bare reflection coefficients, such a quantum crossover can be achieved by scanning the conductance profile over the gate voltage $N_g$. 

By adding more ballistic channels connecting two islands and turning the left and right QPCs to a fully ballistic regime, physics of an impurity in LLs with $1/2<\eta<1$ can be realized. In this case, however, the fully ballistic left and right electronic channels will completely destroy the interesting physics associated with resonant tunneling, and consequently the quantum critical point and Kondo effect will be absent.

In closing, we showed that a nanoelectronic circuit comprising hybrid metal-semiconductor islands coupled to each other by a set of ballistic electronic channels can provide a promising route for quantum simulating exotic zero temperature quantum critical phenomena associated with Luttinger liquid physics. We discussed the realization of Kondo effects embedded in an LL environment using only the well-defined quantum Hall channels. The explored quantum critical points in proposed hybrid setups characterized by fractionalized excitations could offer a coveted pathway for further exploring anyons, given their broader interest in quantum computation. We also demonstrated that when some of the connecting constrictions deviate from the full ballistics, the couple hybrid system flows toward the intermediate coupling fixed point characterized by the oscillatory behavior of the corresponding transport properties.

We are grateful to K. A. Matveev for many fruitful discussions and for comments on an earlier
version of this draft. We are thankful to Christophe Mora for previous collaborations on related projects. We also thank Mikhail Kiselev for useful discussions. This work is supported by the U.S. Department of Energy, Office of Science, Basic Energy Sciences, Materials Sciences and Engineering Division.
\bibliography{biblio}
\end{document}